\documentclass{article}

\usepackage{PRIMEarxiv}

\usepackage[utf8]{inputenc} 
\usepackage[T1]{fontenc}    
\usepackage{hyperref}       
\usepackage{url}            
\usepackage{booktabs}       
\usepackage{amsfonts}       
\usepackage{nicefrac}       
\usepackage{microtype}      
\usepackage{lipsum}
\usepackage{fancyhdr}       
\usepackage{graphicx}       
\graphicspath{{media/}}     
\usepackage{times}
\usepackage[authoryear]{natbib}
\usepackage{amsmath,amssymb,amsthm,amscd,verbatim, graphicx, geometry, colortbl, xcolor, natbib, bm, float}
\usepackage{thmtools}
\usepackage{thm-restate}
\usepackage{appendix}
\usepackage{textcomp}
\usepackage{booktabs}
\usepackage{multirow}
\usepackage{courier}
\usepackage{url}
\usepackage[flushleft]{threeparttable}
\usepackage{xcolor}
\usepackage{hyperref}
\newtheorem{theorem}{Theorem}[section]

\title{Extending the coefficient of variation for measuring heterogeneity following a meta-regression
}

\author{
  Maxwell Cairns, Luke A. Prendergast \\
  Department of Mathematics and Statistics \\
  La Trobe University \\
  Melbourne, Australia\\
  \texttt{mrcairns994@gmail.com, luke.prendergast@latrobe.edu.au} \\
}

\begin{document}
\maketitle

\begin{abstract}
Meta-regression is often used to form hypotheses about what is associated with heterogeneity in a meta-analysis and to estimate the extent to which effects can vary between cohorts and other distinguishing factors. However, study-level variables, called moderators, that are available and used in the meta-regression analysis will rarely explain all of the heterogeneity.  Therefore, measuring and trying to understand residual heterogeneity is still important in a meta-regression, although it is not clear how some heterogeneity measures should be used in the meta-regression context.  The coefficient of variation, and its variants, are useful measures of relative heterogeneity. We consider these measures in the context of meta-regression which allows researchers to investigate heterogeneity at different levels of the moderator and also average relative heterogeneity overall. We also provide CIs for the measures and our simulation studies show that these intervals have good coverage properties. We recommend that these measures and corresponding intervals could provide useful insights into moderators that may be contributing to the presence of heterogeneity in a meta-analysis and lead to a better understanding of estimated mean effects.
\end{abstract}

\section{Introduction}
In meta-analyses, the presence of heterogeneity is common and is seen by many as a nuisance that contributes to estimator variability. However, if we can identify one or more moderating variables, called moderators, that can account for at least some of the heterogeneity, then heterogeneity may give valuable insights. A successful moderator analysis can identify variables, the independent variables (IVs) or moderators, that vary over studies and contribute to variation in the dependent variable (DV) which is the effect measure of interest.  No single study may have manipulated a particular variable as an IV, but if that variable varies over studies, and is identified as a moderator, then the meta-analysis is giving fresh insight that may guide future research. 

A very useful type of moderator analysis is meta-regression. A meta-regression is analogous to linear regression where we model the DV against moderators of interest. For example, \cite{de2009standard} uses the standard quality of participant care as a moderating variable. 

There are two common types of moderators: numerical moderators and factor moderators. A numerical moderator simply takes numerical values for $x_i$ for the $i$-th study and $\beta_1$ is interpreted as the change in the true mean effect for each unit change of $x_i$; for example, the data considered by \cite{colditz1994efficacy} which considers the effect of a tuberculosis vaccine.  For this meta-analysis a numeric moderator is absolute latitude as a geographical indicator of where a study was conducted. A factor moderator, either ordinal or nominal, takes values for different levels of the moderator. In \cite{colditz1994efficacy},  allocation method (random, alternate or systemic) is a factor moderator indicating how treatment was allocated for each study.  In general, when we are estimating effects we are interested in moderators which measure sample characteristics (e.g., average or median cohort age). We would not be interested in moderators such as source characteristics (e.g., researcher characteristics) although the question of whether this helps explain some heterogeneity may still be relevant. \cite{card2015applied} provides a good discussion of the different moderator types, which may provide useful guidance in regards to identifying potential moderators of interest.  Additional information and advice on conducting meta-regression analyses is provided in \cite{borenstein2011introduction, baker2009understanding, higgins2004controlling, thompson2002should}.

While meta-regression can explain some heterogeneity, it is often the case that there is still residual heterogeneity remaining.  Therefore, studying heterogeneity following a meta-regression is still important and we believe that a useful measure of heterogeneity to complement meta-regression is the coefficient of variation (CV) \citep[][]{takkouche1999evaluation,takkouche2013confidence}.  Here, the CV is the the between-studies standard deviation divided by the absolute effect.  A small CV suggests that, while heterogeneity may be present, we do not expect the effects to vary much relative to their magnitude.  On the other hand, a large CV tells us that the effects can be noticeably different, and even contradictory to one another (e.g., a negative effect in some cases, and positive in others).  In this paper we consider using the CV in the context of meta-regression.  This can allow us to obtain a deeper understanding of heterogeneity, and explore questions such as \textit{`at which geographical locations is the vaccine effect likely to be most variable and therefore less reliable?'} and \textit{`what is the average heterogeneity in this meta-analysis?'}. 

The structure of the rest of this paper is as follows.  In Section 2 we provide background material on the meta-regression model, inverse variance weights estimation and estimators of the heterogeneity variance.  In Section 3 we discuss the CV as a measure of heterogeneity.  In Section 4 we consider mean CV and associated variances before considering confidence intervals in Section 5.  Simulations are provided in Section 6 that illustrate very good coverage of our preferred intervals and examples are also considered before the conclusion in Section 7.

\section{Meta-analysis and meta-regression analysis}

In this section we briefly introduce background material on the meta-regression model, inverse variance weights estimation and estimators of the heterogeneity variance.

\subsection{The models}

Let $Y_1,\ldots,Y_k$ be estimators of an effect for $k$ independent studies.  In the presence of zero heterogeneity, the fixed effect model (FEM) is
$$Y_i=\beta + \epsilon_i,\;\;i=1,\ldots,k,$$
where $\epsilon_i\sim N(0,v_i)$ is the sampling error and $v_i$ is the variance of the estimator $Y_i$ for the $i$-th study.  The FEM is rarely justified, since differences in studies (e.g., treatment dosage, population demographics etc.) can reasonably be assumed to result in differences of the effects to be estimated.  These differences in the effects are simply referred to as \textit{heterogeneity}.  To allow for this heterogeneity, we can define the random effects model (REM) as
\begin{equation}
    Y_i = \beta + \gamma_i + \epsilon_i,
    \label{Eq:rem}
\end{equation}
where $\gamma_i \sim N(0, \tau^2)$ is the random effect that allows effects to vary due to unexplained heterogeneity.

A simple meta-regression is modelled in a similar way to simple linear regression. That is, we have an intercept ($\beta_0$), a slope coefficient ($\beta_1$) and $x_1,\ldots,x_k$ are observed moderator values for the $k$ studies.  The simple REM meta-regression model is
\begin{equation}
    Y_i = \beta_0 + \beta_1\cdot x_i + \gamma_i + \epsilon_i,
    \label{Eq:regMod}
\end{equation}
where $\beta_0=E(Y|x=0)$ is the mean effect for an arbitrary study when the moderator is equal to zero.  More generally,  $E(Y|x)=\beta_0 + \beta_1\cdot x$ and given estimates of $\beta_0$ and $\beta_1$ we can estimate the mean effects for given values of the moderator. 

\subsection{Inverse variance weight estimation}

A pooled effect estimator of $\beta$ in the FEM and REM in the form of a weighted average of the $Y_1,\ldots,Y_k$ is $$\widehat{\beta}=w_1Y_1+\ldots+w_kY_k,$$ where $w_1,\ldots,w_k$ are weights that sum to one.  Note that $E(\widehat{\beta})=\beta$ where under the FEM, $\beta$ is the common effect over all studies and the mean effect under the REM.

Let $W_i=1/\text{Var}(Y_i)$, the inverse of the variance for the effect estimator for the $i$-th study.  The inverse variance weights (IVW) estimator uses weights that minimises $\text{Var}(\widehat{\beta})$ and these weights are given by $w_i=W_i/\sum^k_{i=1}W_i$. Under the REM in \eqref{Eq:rem}, we have $\text{Var}(Y_i)=v_i+\tau^2$ and so $W_i=1/(v_i+\tau^2)$.  In both cases the IVW estimator and its variance are
\begin{equation}\label{theta_hat}
\widehat{\beta} = \frac{\sum^k_{i=1}W_iY_i}{\sum^k_{i=1}W_i}=\sum^k_{i=1}w_iY_i,\;\;\text{Var}(\widehat{\beta})=\sum^k_{i=1}\frac{1}{W_i}.
\end{equation}

Let $\bm{\beta}=[\beta_0,\beta_1]^\top$, $\mathbf{W}=\text{diag}(W_1,\ldots,W_k)$ be a diagonal matrix of the un-scaled weights, $\mathbf{X}$ be the matrix of moderators where the $i$-th row is $[1, x_i]$ and $\mathbf{Y}=[Y_1,\ldots,Y_k]$.  Then the pooled estimator of $\bm{\beta}$ for the meta-regression model is the weighted least squares estimator where
\begin{equation}\label{betahat}
    \widehat{\bm{\beta}}=\left[\begin{array}{c}
         \widehat{\beta}_0  \\
          \widehat{\beta}_1 
    \end{array}\right]=\left(\mathbf{X}^\top \mathbf{W} \mathbf{X}\right)^{-1}\mathbf{X}^\top \mathbf{W}\mathbf{Y},\;\text{Var}(\widehat{\bm{\beta}})=\left[\begin{array}{cc}
         \text{Var}(\widehat{\beta}_0) & \text{Cov}(\widehat{\beta}_0,\widehat{\beta}_1) \\
         \text{Cov}(\widehat{\beta}_0,\widehat{\beta}_1)  & \text{Var}(\widehat{\beta}_1)
    \end{array}\right]=\left(\mathbf{X}^\top \mathbf{W} \mathbf{X}\right)^{-1}.
\end{equation}

Note that the weights for the REM depend on $\tau^2$, the variance of the random effect.  In practice this is unknown and needs to be estimated.  Once estimated, $\tau^2$ is replaced by its estimator $\widehat{\tau}^2$ in the above formula.  We next discuss some measures of $\tau^2$. 

\subsection{Estimators of heterogeneity variance}

There are many estimators of $\tau^2$, including maximum-likelihood and restricted maximum-likelihood approaches, as well as methods of moments and unbiased methods. Here, we will describe two methods. Discussion of other approaches can be found in \cite{veroniki2016methods,langan2019comparison}.

The most commonly used method is the DL measure proposed by \cite{dersimonian1986meta}. Let $SF_j = \sum_{i = 1}^k(W_{i, FE})^j$ where $W_{i,FE}$ is the weight assigned to the $i$-th study of a fixed-effect (FE) meta-analysis. The authors define
\begin{equation}
\widehat{\tau}^2 = \text{max}\left\lbrace0, \frac{Q - (k - 1)}{SF_1} - \frac{SF_2}{SF_1}\right\rbrace,
\label{hetDL}
\end{equation}
where $Q = \sum W_{i, FE}(Y_i - \widehat{\beta}_{FE})^2$ and $\widehat{\beta}_{FE}$ is the meta-estimate of the FE analysis. This method is commonly referred to as the DerSimonian and Laird (DL) method. 

A second estimator that we consider is the REML estimator. \cite{veroniki2016methods,langan2019comparison} recommend the REML estimator as an estimator for $\tau^2$, however it is not available in all analysis packages. Further details about the REML estimator can be found in \cite{sidik2005simple}.
Throughout this paper we use the DL and REML estimators of $\tau^2$. It is important to note however, that the measures will work with any estimator of $\tau^2$.

\section{The coefficient of variation}\label{ch4measures}

A very common measure of uncertainty in a meta-analysis is $I^2$ \citep{higgins2002quantifying} defined to be 
\begin{equation*}
    I^2 = \frac{\tau^2}{\tau^2 + \sigma_Y^2},
\end{equation*}
where $\sigma_Y^2$ is a typical within-studies variance.  However,  $I^2$ can be small even when heterogeneity is large (i.e. this can happen when a typical sampling error variance is large) or large when heterogeneity is small (such as when sampling errors are small).  For more on this see \cite{borenstein2017basics}.  Hence, it is important that $I^2$ is interpreted correctly as a measure of heterogeneity relative to total variability in the specific meta-analysis.   

A measure which may be very useful for quantifying heterogeneity is the coefficient of variation (CV)\citep[][]{takkouche1999evaluation, takkouche2013confidence} defined to be
\begin{equation*}
    \text{CV}_B = \frac{\tau}{|\beta|}.
\end{equation*}
A large CV$_B$ indicates that heterogeneity is large relative to the magnitude of the effect.  When this occurs the effects can even be contradictory, depending on, e.g., cohorts differences such as age and gender, conditions under which a treatment is implemented etc.  On the other hand, a small CV$_B$ indicates that we have a much more accurate understanding of the magnitude of the effect, even when noting such differences.  Hence, the coefficient of variation can be a useful tool to help describe heterogeneity, perhaps as a complement to other measures such as $I^2$ \citep{higgins2002quantifying}.  However, when $\beta$ is very small, the CV can be huge and such cases there may be reluctance to use it.  Additionally, given the prominent use of $I^2$, that is on a simple-to-understand scale, not everyone may be comfortable reporting CV$_B$.  Therefore, \cite{cairns2020ratio} transformed the CV as 
\begin{equation*}
    M_1 = \frac{\tau}{\tau + |\beta|} \text{ and } M_2 = \frac{\tau^2}{\tau^2 + \beta^2}
\end{equation*}
and provided several confidence intervals (CIs) that had good coverage properties in a wide variety of scenarios. In that paper they discussed several properties which we believe could be important for measures of heterogeneity. Importantly, the paper noted that we can directly link $\text{CV}_B$, $M_1$ and $M_2$ in the following ways:
\begin{equation}
    \log(\text{CV}_B)=\text{logit}(M_1)=\text{logit}(M_2)/2,
\label{E:CVlinkch4}
\end{equation}
and 
\begin{equation}
    M_1=\frac{\text{CV}_B}{1 + \text{CV}_B}\;\;\text{and}\;\;M_2=\frac{\text{CV}^2_B}{1 + \text{CV}^2_B}.
    \label{E:CVlink2ch4}
\end{equation}

\section{CV estimation in the context of meta-regression}\label{ch4means}

In this section we will briefly discuss the CV in the context of meta-regression. For a given moderator value $x$, we define $\beta_x = \beta_0 + \beta_1x$ as the true value to be estimated.  Note that we can similarly consider more than one moderator variable, such as in multiple linear regression, including factor moderators that require the $x$ (or multiple $x$-s) to be dummy variables indicating factor level. Then, $\text{CV}_B=\frac{\tau}{|\beta_x|}$ and we can estimate the CV by
\begin{equation}
    \widehat{\text{CV}}_B=\frac{\widehat{\tau}}{|\widehat{\beta}_x|}.\label{CV_hat}
\end{equation}
Note that if changes to a moderator can effect significant change to the mean effect, then very different values for the CV may be result for different $x$-s.   For example, consider two scenarios:

\begin{description}
\item[Small $\widehat{\text{CV}}_B$] since we estimate that $\tau$ is small relative to $\beta_x$, then for this $x$ we have a good understanding of the magnitude of the effect (e.g., we estimate that the magnitude of $\beta_x$ will not vary much for studies conducted for this value of $x$).

\item[Large $\widehat{\text{CV}}_B$] since we estimate that $\tau$ is large relative to $\beta_x$, then for this $x$ we have a poor understanding of the magnitude of the effect (e.g., we estimate that the magnitude of $\beta_x$ will vary a lot for studies conducted for this value of $x$).
\end{description}

Note that in the above we refer to the fact that we are only estimating relative heterogeneity and our conclusions need to reflect that.  Hence, to provide some rigour we need CIs which we consider in the next section.  Variances for estimators of $\tau$ are readily available, and we can easily obtain a variance for $\widehat{\beta}_x$ as
\begin{equation}
    \text{Var}(\widehat{\beta}_x) = \text{Var}(\widehat{\beta}_0 + x\widehat{\beta}_1) = \text{Var}(\widehat{\beta}_0) + x^2\cdot \text{Var}(\widehat{\beta}_1) + 2\cdot x \cdot \text{Cov}(\widehat{\beta}_0, \widehat{\beta}_1),
    \label{varBeta}
\end{equation}
where the variances and covariances are given in \eqref{betahat}.  

\cite{takkouche2013confidence} considered Wald CIs for the log of CV$_B$.  For the estimator of $M_1$ (and $M_2$), \cite{cairns2020ratio} showed that improved intervals were obtained when these were transformed using the logit transformation and some different interval methods to be discussed later.  Hence, using their variance of the logit transformed estimation and also the link between this and the log of the CV estimator shown in \eqref{E:CVlinkch4},  we have
\begin{equation*}
    \text{Var}\left[\text{log}(\widehat{\text{CV}}_B)\right]=\text{Var}\left[\text{logit}(\widehat{M}_1)\right]\approx \text{Var}(\widehat{\tau}^2)\cdot \frac{1}{4\tau^4} + \text{Var}(\widehat{\beta_x})\cdot \frac{1}{\beta_x^2}.
\end{equation*}

Note in the above we have considered a single fixed $x$, but we may wish to explore different $x$'s.  For example, if we assume that $\text{CV}_B$ has been calculated for each level of a moderator, then one overall measure is an average of these $\text{CV}_B$'s. We now consider several possibilities: a weighted average of $\log$ values, a weighted average and a geometric mean. For these averages we focus directly on the CV$_B$, however the average of log values can also be considered as the average logit of estimators of $M_1$.

Let $\bm{x} = [x_1,\ldots,x_r]^\top$ be a vector of moderator values. This can either be chosen by the researcher, or they may be the specific moderator values observed for the $k$ studies (in which case $r = k$). Further, let $\bm{\omega} = [\omega_1,\ldots, \omega_r]^\top$ be weights that will be assigned to each $x_r$ for the purpose of weighted average of the heterogeneity measures computed at these moderator values. Finally, let $\text{CV}_{B, i}$ be value of $\text{CV}_B$ calculated for each $x_i$. Then the weighted average of $\log(\text{CV}_B)$ estimates and the corresponding geometric mean are of the form
\begin{equation}
   \widehat{\text{waLCV}} = \bm{\omega}^\top
    \cdot
    \log\left(\begin{bmatrix}
        \widehat{\text{CV}}_{B, 1}\\
        \cdot \\
        \cdot \\
        \cdot \\
        \widehat{\text{CV}}_{B,r}
    \end{bmatrix}\right),\;\; \widehat{\text{GM}}=\exp(\widehat{\text{waLCV}}),
    \label{E:lCVreg}
\end{equation}
where $\omega_1 = \dots = \omega_r = 1/r$. Finally, we define the weighted average of CV$_B$'s as 
\begin{equation}
    \widehat{\text{waCV}} = \bm{\omega}^\top
    \cdot
    \begin{bmatrix}
        \widehat{\text{CV}}_{B, 1}\\
        \cdot \\
        \cdot \\
        \cdot \\
        \widehat{\text{CV}}_{B,r}
    \end{bmatrix}.
    \label{E:waCV}
\end{equation}

Let $\bm{\widehat{\beta}}_X$ be a column vector of estimated values of the $\beta_x$'s corresponding to the $x_i$'s (e.g., the $i$-th element is the estimated $\beta_x$ at $x_i$), $\bm{\widehat{\Sigma}}$ be the variance-covariance matrix of $\bm{\widehat{\beta}}_X$, $\text{sign}(\bm{\widehat{\beta}}_X)$ be a column vector of corresponding signs ($\pm1$) of the elements of $\bm{\widehat{\beta}}_X$, $\bm{\widehat{\beta}}^{\downarrow}_X$ be the vector where element $\widehat{\beta}_{i}^{\downarrow} = 1/\widehat{\beta}_{x_i}$ and $|\bm{\widehat{\beta}}^{\downarrow}_X|$ be the vector comprised of elements $|\widehat{\beta}_{x_i}^{\downarrow}|$. Then, we use the following notation: we let $\odot$ be the Hadamard Product (in which matrices are multiplied together element-wise), let $\bm{I}_H$ be the multiplicative identity of the Hadamard product (i.e., a $r \times r$ matrix containing $1$'s in every entry) and $\cdot$ be the usual matrix multiplication. We find the following theorem:
\begin{theorem}\label{thm:varGeo}
An approximate variance-covariance matrix for $r$ dependent log transformed $\text{CV}_B$ estimators is
\begin{equation*}
        \bm{V}_{vlcv} = \left[\frac{\text{Var}(\widehat{\tau}^2)}{4\cdot\widehat{\tau}^4} \cdot \bm{I}_H\right] + \left[(\text{sign}(\bm{\widehat{\beta}}_X)\cdot\text{sign}(\bm{\widehat{\beta}}_X)^\top)\odot(|\bm{\widehat{\beta}}^{\downarrow}_X|\cdot |\widehat{\bm{\beta}}^{\downarrow}_X|^\top)\odot\bm{\widehat{\Sigma}}\right].
\end{equation*}
\end{theorem}
Similarly, we can calculate a variance-covariance matrix for $r$ dependent $\text{CV}_B$'s.
\begin{theorem}\label{thm:varCVREG}
Let $\bm{\widehat{\beta}}^{\downarrow2}$ be the column vector containing elements $(\beta_{i}^{\downarrow})^2$. Then an approximate variance-covariance matrix for $k$ dependent $\text{CV}_B$ estimators is
\begin{align*}
        \bm{V}_{vcv} &= \left[\left(\frac{\text{Var}(\widehat{\tau}^2)}{4\cdot\widehat{\tau}^2} \cdot \bm{I}_H\right)\odot (|\bm{\widehat{\beta}}^{\downarrow}_X|\cdot |\bm{\widehat{\beta}}^{\downarrow}_X|^\top)\right]\\ &+ \left[(\widehat{\tau}^2\cdot\bm{I}_H)\odot(\text{sign}(\bm{\widehat{\beta}}_X)\cdot \text{sign}(\bm{\widehat{\beta}}_X)^\top)\odot(\bm{\widehat{\beta}}^{\downarrow2}_X\cdot\bm{\widehat{\beta}}_X^{{\downarrow2}^{\top}})\odot\bm{\widehat{\Sigma}}\right].
\end{align*}
\end{theorem}
The proofs of Theorems \ref{thm:varGeo} and \ref{thm:varCVREG} can be found in Appendix A and B respectively. 

When performing a factor moderator analysis, the method to choose weights is very important. A factor moderator provides a vector (i.e., $\bm{x}$) containing a value for each level of the moderator. A researcher could then choose to use equal weights across each level of the vector, they could set the weights equal to the proportion of times a value appears in the data set or they could choose the weights based on knowledge of the research area in practice (e.g., giving a small weight to a factor level that is known to represent a very rare occurrence or type). Let us refer to the example seen in the introduction. Looking at the allocation vector in the \cite{colditz1994efficacy} data set, we see there are three levels: random, alternate and systemic. Therefore, if the researcher chooses equal weights we would have $\omega_1 = \omega_2 = \omega_3 = \frac{1}{3}$. If the researcher chose to define the weights as the proportion of the times the level appeared, they would have $\omega_1 \approx 0.538$, $\omega_2 \approx 0.154$ and $\omega_3 = 0.308$. Finally, the researcher may decide that a random allocation is much more likely than what is represented in the data set, and therefore increasing that weight while decreasing others (e.g., choosing $\omega_1 = 0.7$, $\omega_2 = 0.1$ and $\omega_3 = 0.2$).

\section{Confidence intervals}\label{ch4CI}

\cite{cairns2020ratio} proposed four methods of calculating CIs in the context of meta-analysis. These methods naturally extend to the meta-regression framework. Using the \texttt{metafor} \citep{metafor} package in \texttt{R} \citep{R}, we can readily calculate a CI for $\beta_x$. We provide code for this in the OSF page found \href{https://osf.io/f68za/}{here}. We use this method of calculating intervals for $\beta_x$ in all substitution based methods of calculating intervals. In Sections \ref{ch4subPropImp}, \ref{ch4subBoth} and \ref{ch4subWT} we briefly define the intervals we use in this paper.

In the sections below we define CIs for $M_1$. CIs for CV$_B$ and $M_2$ can be found using the transformations seen in Equations \ref{E:CVlinkch4} and \ref{E:CVlink2ch4}.

\subsection{PropImp}\label{ch4subPropImp}

Let $f(X, Y)$ be a function that is monotonic in both $X$ and $Y$, where $(X_{z(\alpha)}^L, X_{z(\alpha)}^U)$ is a $(1 - \alpha)\times100\%$ CI for $X$ and $z(\alpha)=\Phi^{-1}(1-\alpha/2)$ where $\Phi^{-1}$ is the standard normal inverse distribution function; e.g.,  $\Phi^{-1}(0.975)\approx 1.96$. The CI for $Y$,  $(Y_{z(\alpha)}^L, Y_{z(\alpha)}^U)$, is similarly defined. Then the PropImp interval \citep{newcombe2011propagating} for $M_1$ would have the form
\begin{equation*}
    L = \min_{0 \leq \theta_1 \leq \pi/2}\frac{\widehat{\tau}_{z\sin(\theta_1)}^L}{\widehat{\tau}_{z\sin(\theta_1)}^L + |\widehat{\beta}_{z\cos(\theta_1)}|^U},\ U = \max_{0 \leq \theta_2 \leq \pi/2}\frac{\widehat{\tau}_{z\sin(\theta_2)}^U}{\widehat{\tau}_{z\sin(\theta_2)}^U + |\widehat{\beta}_{z\cos(\theta_2)}|^L},
\end{equation*}
remembering that $\beta$ is referring to $\beta_x = \beta_0 + \beta_1x$. The full PropImp algorithm for the CV is discussed in \cite{cairns2020ratio}.

\subsection{Alpha adjusted intervals}\label{ch4subBoth}

Similarly, let $[L_\tau(\alpha), U_\tau(\alpha)]$ and $[L_{\beta_x}(\alpha), U_{\beta_x}(\alpha)]$ denote $(1-\alpha)\times 100$\% CIs for $\tau$ and $\beta_x$ respectively. Then when using the method prescribed in \cite{newcombe2011propagating, lloyd1990confidence} for calculating CIs using CIs for both variables, we see intervals for $M_1$ of the form 
\begin{equation}
        \left[\frac{L_\tau(\alpha)}{L_\tau(\alpha) + U_{|\beta_x|}(\alpha)},\ \frac{U_\tau(\alpha)}{U_\tau(\alpha) + L_{|\beta_x|}(\alpha)}\right].
        \label{E:lloyd}
\end{equation}

This method of calculating CIs works by assigning $\alpha = 0.8342$ to make intervals less conservative. As such, we are using the same formula seen in Equation \ref{E:lloyd} but are calculating a CI(0.8342, 0.8342).

\subsection{Wald-type intervals}\label{ch4subWT}
A Wald-type (WT) CI is a possible alternative to the substitution methods we describe above. Here we consider 95\% CIs and therefore choose $z_{1-\alpha/2}=z_{0.975}=1.96$ as our pivot, however other values can be used to achieve different levels of confidence.  Therefore a WT interval for the logit of $M_1$ in the context of meta-regression has the form
\begin{equation*}
    \text{Logit}(\widehat{M_1}) \pm1.96 \sqrt{\text{Var}\left[\text{Logit}(\widehat{M_1})\right]},
\end{equation*}
with variance
\begin{equation*}
    \text{Var}\left[\text{logit}(\widehat{M}_1)\right]\approx \text{Var}(\widehat{\tau}^2)\cdot \frac{1}{4\tau^4} + \text{Var}(\widehat{\beta_x})\cdot \frac{1}{\beta_x^2},
\end{equation*}
where Var$(\widehat{\beta_x})$ is from \eqref{varBeta}. We then back transform to the regular scale using the usual methods.  Additionally, given the link between the logit of $M_1$ and the log of CV$_B$ given in \eqref{E:CVlinkch4}, the above is also a confidence for $\log (\text{CV}_{B})$.

\subsection{Confidence intervals for mean values}\label{ch4submeansCI}

CIs of mean values may also be of interest. Here we consider CIs for the case where we take a weighted or geometric mean of the estimated $\beta_i$'s. These intervals are regular WT intervals of their respective values and the required variance matrices can be found in Theorems \ref{thm:varGeo} and \ref{thm:varCVREG}. A weighted average of $\log(\text{CV}_B)$'s would have the interval
\begin{equation}
    \widehat{\text{waLCV}} \pm 1.96\sqrt{\bm{\omega}^\top\cdot\bm{V}_{vlcv}\cdot\bm{\omega}}.
    \label{regWTlogCI}
\end{equation}
A CI for a geometric mean of $\text{CV}_B$'s has the form
\begin{equation}
    \text{exp}\left\lbrace \widehat{\text{waLCV}} \pm 1.96\sqrt{\bm{\omega}^\top\cdot\bm{V}_{vlcv}\cdot\bm{\omega}}\right\rbrace,
    \label{regWTgeoCI}
\end{equation}
where we assume equal weights (i.e., $\omega_1 = \ldots = \omega_r = 1/r$).  A CI for a weighted average of $\text{CV}_B$'s is defined as follows:
\begin{equation}
    \widehat{\text{waCV}} \pm1.96 \sqrt{\bm{\omega}^\top\cdot\bm{V}_{vcv}\cdot\bm{\omega}},
    \label{regWTCI}
\end{equation}
where $\bm{V}_{vcv}$ is the variance-covariance matrix defined in Theorem \ref{thm:varCVREG}. 

\section{Simulations and examples}\label{ch4sims}

In this section we display the results of several simulation studies. These studies were performed using the \texttt{metafor} package \citep{metafor} in \texttt{R} \citep{R}. We simulate our data based on the settings from real data sets. We randomly generate our $Y_i$'s from a normal distribution with mean $\beta_0 + \beta_1x_i$ and variance $\tau^2 + v_i$ (remembering that $v_i$ is the within-study variance for the $i$-th study). We will now briefly describe the data sets used to create our simulations. The Studies of the Effectiveness of the BCG Vaccine Against Tuberculosis (BCG) data set \citep{colditz1994efficacy} contains 13 studies. We performed simulations using the numeric moderator absolute latitude and a second set of simulations using the factor moderator allocation.  The next data set, also containing 13 studies, was the Standard Care Quality (SCQ) and HAART-adherence (HAART) data set \citep{de2009standard}. For simulations based on this data set we used the standard quality of care as a moderating variable. We also performed simulations using the factor moderator ethnicity as a moderating variable.  Due to the size of the data sets we do not describe them fully here. However they can be viewed using the code provided on our OSF page.

\subsection{Simulation results}\label{ch4subres}

We will now discuss the results of the above simulations.

\subsubsection{Numeric moderator}\label{ch4subsubNumRes}

We begin by providing the results of the simulations using numeric moderators. For simplicity, when computing intervals for a specified moderator value, we have set the moderator values equal to those observed.  However, it is of course possible to choose any value for the moderators. 
   
    \begin{table}[h!t]
        \centering
        \caption{Coverages calculated using 10,000 trials with moderator values chosen to be the absolute latitude values found in the \cite{colditz1994efficacy} data set.}
        \bigskip
        \resizebox{\columnwidth}{!}{
        \begin{tabular}{l|ccccccccc}
        \hline
           Mods & 13 & 18 & 19 & 27 & 33 & 42 & 44 & 52 & 55 \\
           \hline
            $\alpha_{adj}$ & 0.937 & 0.952 & 0.951 & 0.937 & 0.922 & 0.914 & 0.914 & 0.916 & 0.917 \\ 
            $\alpha_{adj}$ [REML] & 0.936 & 0.957 & 0.956 & 0.946 & 0.931 & 0.923 & 0.922 & 0.926 & 0.927 \\ 
            Prop & 0.970 & 0.977 & 0.977 & 0.972 & 0.967 & 0.964 & 0.964 & 0.965 & 0.966 \\ 
            Prop [REML] & 0.971 & 0.980 & 0.980 & 0.977 & 0.973 & 0.970 & 0.969 & 0.970 & 0.970 \\ 
            WT & 0.939 & 0.987 & 0.990 & 0.999 & 0.997 & 0.995 & 0.995 & 0.995 & 0.995 \\ 
            WT [REML] & 0.921 & 0.978 & 0.982 & 0.997 & 0.992 & 0.987 & 0.987 & 0.988 & 0.989 \\ 
            \hline
        \end{tabular}}
        \label{tab:reg_bcg}
    \end{table}
   
    In Table \ref{tab:reg_bcg} we see coverages for four types of CI for $M_1$ for simulations with absolute latitude as a moderating variable based on the \cite{colditz1994efficacy} example. In general, coverage is strong for each method and typically slightly more conservative when using the REML estimator of $\tau$. We see that coverage for the $\alpha$-adjusted intervals appears to be slightly liberal. The PropImp intervals provide strong coverage for each value of the moderator and the intervals based on the REML estimator tend to be slightly more conservative than the intervals calculated with the DL estimator. This is potentially due to the negative bias shown in the DL estimator when $\tau^2$ is large \citep{veroniki2016methods}. 
    
        \begin{table}[h!t]
        \centering
         \caption{Average widths calculated using 10,000 trials with absolute latitude as a moderating variable. Note that the widths for CV$_B$ based intervals and WT intervals are calculated using medians.}
         \bigskip
         \resizebox{\columnwidth}{!}{
        \begin{tabular}{l|ccccccccc}
        \hline
           Mods & 13 & 18 & 19 & 27 & 33 & 42 & 44 & 52 & 55 \\
           \hline
            $\alpha_{adj}$ ($\text{CV}_B$) & $>1000$ & 7.328 & 4.621 & 1.098 & 0.703 & 0.486 & 0.460 & 0.384 & 0.363 \\
            $\alpha_{adj}$ [REML] ($\text{CV}_B$) &  $>1000$ & 8.840 & 5.318 & 1.161 & 0.741 & 0.512 & 0.485 & 0.406 & 0.384 \\ 
            $\alpha_{adj}$ ($M_1$) & 0.735 & 0.651 & 0.625 & 0.417 & 0.336 & 0.279 & 0.271 & 0.247 & 0.240 \\ 
              $\alpha_{adj}$ [REML] ($M_1$) & 0.718 & 0.642 & 0.619 & 0.418 & 0.338 & 0.283 & 0.276 & 0.254 & 0.247 \\ 
            Prop ($\text{CV}_B$) & $>1000$ & $>1000$ & 18.822 & 1.406 & 0.852 & 0.591 & 0.559 & 0.465 & 0.440 \\ 
              Prop [REML] ($\text{CV}_B$) &  $>1000$ &  $>1000$ & 43.359 & 1.512 & 0.902 & 0.620 & 0.587 & 0.493 & 0.468 \\ 
            Prop ($M_1$) & 0.814 & 0.754 & 0.731 & 0.496 & 0.397 & 0.332 & 0.323 & 0.295 & 0.287 \\ 
              Prop [REML] ($M_1$) & 0.798 & 0.743 & 0.723 & 0.499 & 0.399 & 0.335 & 0.327 & 0.302 & 0.295 \\ 
                WT (CV$_B$) & 0.874 & 0.649 & 0.608 & 0.420 & 0.359 & 0.309 & 0.300 & 0.271 & 0.261 \\ 
                    WT [REML] (CV$_B$) & 0.860 & 0.634 & 0.591 & 0.393 & 0.332 & 0.283 & 0.276 & 0.251 & 0.243 \\ 
  WT ($M_1$) & 0.874 & 0.649 & 0.608 & 0.420 & 0.359 & 0.309 & 0.300 & 0.271 & 0.261 \\ 
  WT [REML] ($M_1$) & 0.860 & 0.634 & 0.591 & 0.393 & 0.332 & 0.283 & 0.276 & 0.251 & 0.243 \\ 
            \hline
        \end{tabular}}
        \label{tab:reg_bcg_width}
    \end{table}

In Table \ref{tab:reg_bcg_width} we show the average widths corresponding to the coverages seen in Table \ref{tab:reg_bcg}. We show results for CV$_B$, $M_1$ and $M_2$. We note that particularly for small $x_i$-s the CV$_B$ intervals are very wide. However, this is because the CV$_B$ values are all extremely large, e.g., from 10000 trials the average $\widehat{\text{CV}}_B\approx 13$ when $x = 13$ and in this situation estimates of the transformed versions $M_1$ and $M_2$ may be preferred.  
    
       \begin{table}[h!t]
        \centering
        \caption{Coverages calculated using 10,000 trials with moderator values chosen to be the standard quality of care values found in the \cite{de2009standard} data set.}
        \bigskip
        \resizebox{\columnwidth}{!}{
        \begin{tabular}{l|ccccccccccccc}
        \hline
           Mods & 3.67 & 4.80 & 7.47 & 7.80 & 8.60 & 10.47 & 12.07 & 15.07 & 20.00 & 21.00 & 22.67 & 22.80 & 27.80 \\
           \hline
            $\alpha_{adj}$ & 0.951 & 0.950 & 0.944 & 0.943 & 0.941 & 0.937 & 0.933 & 0.928 & 0.934 & 0.935 & 0.937 & 0.937 & 0.942 \\
            $\alpha_{adj}$ [REML] & 0.953 & 0.953 & 0.948 & 0.947 & 0.946 & 0.943 & 0.939 & 0.932 & 0.938 & 0.940 & 0.942 & 0.942 & 0.946 \\ 
            Prop & 0.974 & 0.973 & 0.973 & 0.972 & 0.972 & 0.973 & 0.973 & 0.971 & 0.972 & 0.972 & 0.973 & 0.973 & 0.972 \\ 
              Prop [REML] & 0.973 & 0.973 & 0.974 & 0.974 & 0.973 & 0.974 & 0.975 & 0.974 & 0.974 & 0.974 & 0.974 & 0.974 & 0.973 \\ 
              WT & 0.966 & 0.967 & 0.969 & 0.969 & 0.969 & 0.969 & 0.969 & 0.968 & 0.968 & 0.968 & 0.968 & 0.968 & 0.966 \\ 
             WT [REML] & 0.961 & 0.961 & 0.962 & 0.963 & 0.964 & 0.965 & 0.964 & 0.963 & 0.964 & 0.964 & 0.963 & 0.963 & 0.961 \\ 
            \hline
        \end{tabular}}
        \label{tab:reg_HAART}
    \end{table}

    \begin{table}[h!t]
        \centering
        \caption{Average widths calculated from 10,000 trials using SCQ as a moderator. The data was simulated from the \cite{de2009standard} data set. Note that the widths for CV$_B$ based intervals and WT intervals are calculated using medians.}
        \bigskip
        \resizebox{\columnwidth}{!}{
        \begin{tabular}{l|ccccccccccccc}
        \hline
           Mods & 3.67 & 4.80 & 7.47 & 7.80 & 8.60 & 10.47 & 12.07 & 15.07 & 20.00 & 21.00 & 22.67 & 22.80 & 27.80 \\
           \hline
           $\alpha_{adj}$ ($\text{CV}_B$) & 0.464 & 0.433 & 0.375 & 0.369 & 0.356 & 0.330 & 0.313 & 0.293 & 0.285 & 0.286 & 0.287 & 0.287 & 0.294 \\ 
           $\alpha_{adj}$ [REML] ($\text{CV}_B$) & 0.510 & 0.476 & 0.410 & 0.403 & 0.388 & 0.358 & 0.339 & 0.317 & 0.310 & 0.311 & 0.314 & 0.314 & 0.326 \\ 
           $\alpha_{adj}$ ($M_1$) & 0.241 & 0.229 & 0.205 & 0.203 & 0.197 & 0.186 & 0.178 & 0.171 & 0.171 & 0.173 & 0.175 & 0.176 & 0.186 \\ 
             $\alpha_{adj}$ [REML] ($M_1$) & 0.253 & 0.240 & 0.214 & 0.212 & 0.205 & 0.193 & 0.186 & 0.178 & 0.180 & 0.181 & 0.185 & 0.185 & 0.198 \\ 
           Prop ($\text{CV}_B$) & 0.582 & 0.534 & 0.448 & 0.439 & 0.420 & 0.386 & 0.367 & 0.346 & 0.333 & 0.334 & 0.337 & 0.337 & 0.357 \\ 
             Prop [REML] ($\text{CV}_B$) & 0.649 & 0.595 & 0.493 & 0.483 & 0.460 & 0.419 & 0.394 & 0.368 & 0.363 & 0.365 & 0.372 & 0.373 & 0.400 \\ 
           Prop ($M_1$) & 0.292 & 0.274 & 0.241 & 0.238 & 0.230 & 0.216 & 0.208 & 0.200 & 0.200 & 0.202 & 0.206 & 0.206 & 0.223 \\ 
             Prop [REML] ($M_1$) & 0.309 & 0.289 & 0.252 & 0.249 & 0.240 & 0.225 & 0.215 & 0.206 & 0.209 & 0.211 & 0.217 & 0.217 & 0.238 \\ 
          WT (CV$_B$)   & 0.246 & 0.239 & 0.224 & 0.222 & 0.219 & 0.212 & 0.207 & 0.200 & 0.195 & 0.194 & 0.194 & 0.194 & 0.194 \\
            WT [REML] (CV$_B$) & 0.236 & 0.227 & 0.210 & 0.209 & 0.205 & 0.197 & 0.192 & 0.186 & 0.182 & 0.183 & 0.183 & 0.183 & 0.186 \\ 
  WT ($M_1$) & 0.246 & 0.239 & 0.224 & 0.222 & 0.219 & 0.212 & 0.207 & 0.200 & 0.195 & 0.194 & 0.194 & 0.194 & 0.194 \\ 
  WT [REML] ($M_1$) & 0.236 & 0.227 & 0.210 & 0.209 & 0.205 & 0.197 & 0.192 & 0.186 & 0.182 & 0.183 & 0.183 & 0.183 & 0.186 \\ 
            \hline
        \end{tabular}}
        \label{tab:reg_deb_widths}
    \end{table}
    
    In Table \ref{tab:reg_HAART} we again see typically good coverage when the settings from \cite{de2009standard} are used.  While the $\alpha$-adjusted coverages are too liberal with the DL, coverage is again slightly more conservative using the REML estimator.  We consider the average widths of the intervals in Table \ref{tab:reg_deb_widths}. We see that the size of the CIs can vary with average widths appearing quite small across the different methods in comparison to the widths seen in Table \ref{tab:reg_bcg_width}.  This is because the estimates to the CV$_B$-s are typically smaller with less relative heterogeneity.  Again we see that the $\alpha$-adjusted intervals are the narrowest. However, narrow interval width comes at the expense of the intervals being too liberal to slightly liberal as we saw in Table \ref{tab:reg_bcg}, in particular when using the DL estimator.  That said, the slightly liberal coverage when using the REML estimator and narrow CIs than the PropImp approach means that the $\alpha$-adjusted intervals with the REML estimator have performed reasonably well.  The slightly wider PropImp intervals align with the slightly conservative coverage seen in Table \ref{tab:reg_HAART}.
    
\begin{figure}[h!t]
    \centering
    \includegraphics[scale = 0.55]{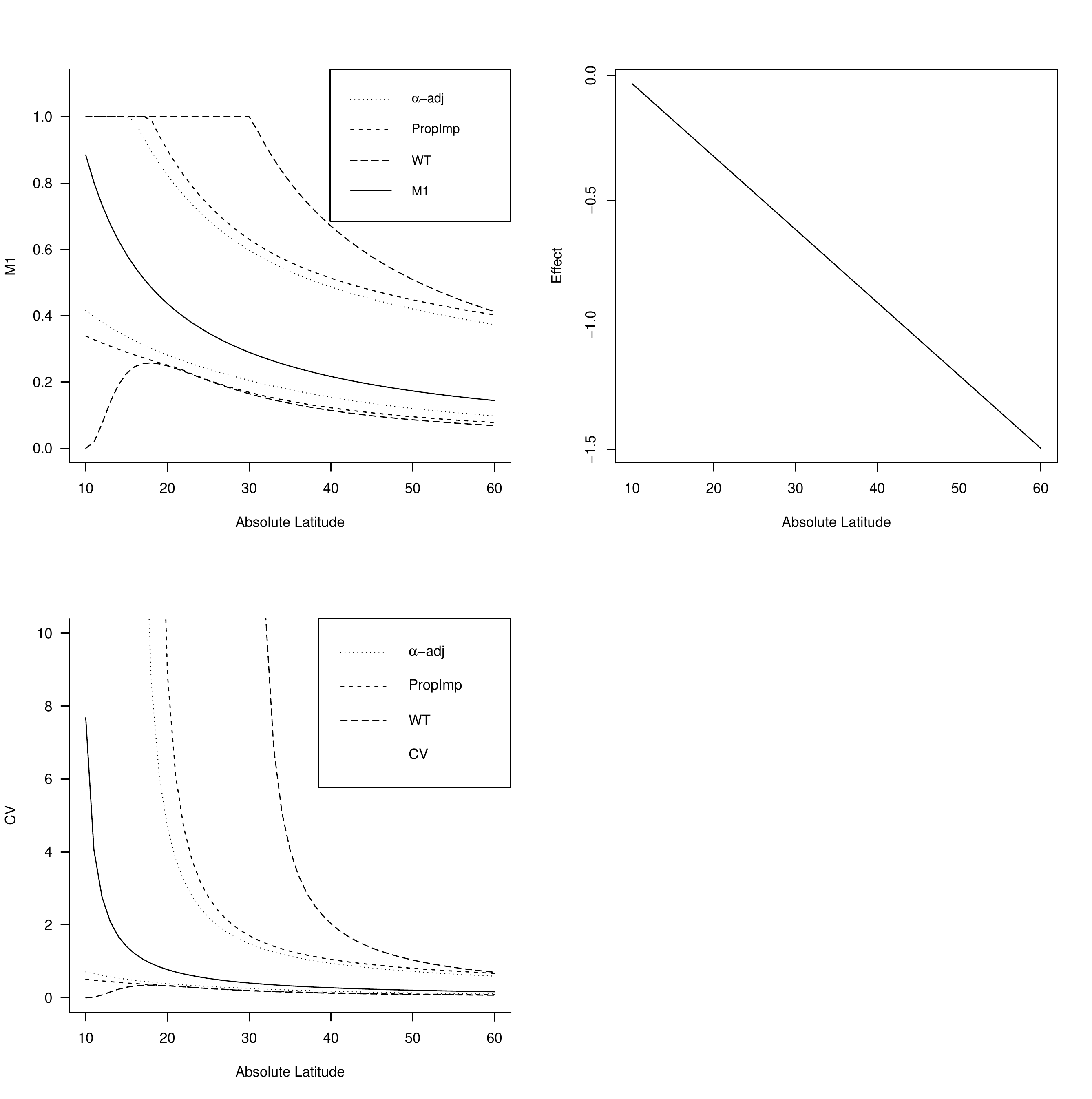}
    \caption{These plots display $M_1$ against absolute latitude (top-left), $\text{CV}_B$ against absolute latitude (bottom-left) as well as several associated CIs and $\widehat{\beta}_x$ against absolute latitude (top-right). These plots highlight the advantage of transforming the CV measures to the $[0, 1]$ scale when the effect ($\beta_x$) is close to zero.}
    \label{fig:reg_scatter}
\end{figure}

In general, we see that the widths of the REML intervals are slightly wider than the DL based intervals but this also corresponds to closer-to-nominal interval estimator.  Hence, when considering both coverage and width, we recommend using the REML estimator.  We also saw that in some examples, huge widths of the intervals were often obtained for CV$_B$ estimation and we noted that this was because of large relative heterogeneity leading to an extremely large upper bound.   Hence, this is one reason why intervals for $M_1$ and $M_2$ may be preferred. To gain a better understanding of this, we see in the top-left plot of Figure \ref{fig:reg_scatter} a plot of the $M_1$ estimate and its $\alpha_{adj}$, PropImp and WT intervals using the DL estimator for $\tau$ (similar results for the REML estimator). In this example, we used the \cite{colditz1994efficacy} data set but defined the moderating variable to be integer values between 10 and 60 inclusive. The WT intervals are the widest (particularly for smaller moderator values) followed by the PropImp intervals, with the $\alpha_{adj}$ intervals providing the shortest intervals. This is what we would expect given our coverages seen in Table \ref{tab:reg_bcg}. The issues with $\text{CV}_B$ being unbounded become obvious in the bottom-left plot in Figure \ref{fig:reg_scatter} where we can see the upper bounds of the CIs are very large. In the top-right plot of Figure \ref{fig:reg_scatter} we see the observed estimated effects against absolute latitude. This helps us highlight the issues with the large values of CV$_B$ for small values of $\widehat{\beta}_x$.  When the estimated effects are close to zero, as is the case for small moderator values, the CV$_B$ measure will become very large and the interval upper bound even larger.  Hence, we can still portray very large relative heterogeneity using $M_1$ (e.g., a large estimate and upper interval bound of one), without the nuisance of reporting extremely large (even infinite) values for CV$_B$.   This is perhaps less problematic for factor moderators which are less likely to have a viable moderator value that results in an estimated effect approximately equal to zero, due to the non-continuous nature of the regression model.

\subsubsection{Factor moderator}\label{ch4subsubFacRes}

In Table \ref{tab:reg_single} we display the results of simulations for $\text{CV}_B$'s as described above. Note that these coverages will be the same for $M_1$ and $M_2$, however the width will differ. 
    
\begin{table}[ht]
\centering
\caption{Coverages and median widths calculated using 10,000 trials with several factor moderators using CV$_B$'s for both DL and REML estimators of $\tau^2$. We show single level data for the two data sets \cite{colditz1994efficacy, de2009standard}.}
\bigskip
\begin{tabular}{llcccc}
  \hline
Data & Level & WT & $\alpha_{adj}$ & PropImp \\ 
  \hline
&Alternate & 0.899 & 0.935 & 0.969\\ 
& Alternate [REML] & 0.911 & 0.944 & 0.971\\ 
 \cite{colditz1994efficacy} & Random & 0.950 & 0.949 & 0.975\\ 
  &Random [REML] & 0.963 & 0.951 & 0.975\\ 
  &Systematic & 0.905 & 0.947 & 0.974\\ 
  &Systematic [REML] & 0.921 & 0.952 & 0.978\\ 
  \hline
  &Other & 0.963 & 0.941 & 0.972\\ 
  &Other [REML] & 0.959 & 0.945 & 0.972\\ 
  \cite{de2009standard} &Caucasian & 0.965 & 0.934 & 0.970 \\ 
  &Caucasian [REML] & 0.961 & 0.938 & 0.972\\ 
   \hline
\end{tabular}
\label{tab:reg_single}
\end{table}

Table \ref{tab:reg_single} results show generally good coverage though the WT intervals can be quite liberal (e.g., 0.899). 

\subsubsection{Geometric and weighted means}\label{ch4subsubmeanRes}

In this section we consider the results of simulations performed for geometric and weighted means.  The $\alpha$-adjusted and PropImp methods no longer have only two substitutions to perform (i.e. the interval for $\tau$ and the interval for the effect) but $r+1$ interval substitutions (one for $\tau$ and then one for each of the moderator values chosen to form the average of the effects where $r=k$ if the observed moderators are chosen).  Hence, we consider the more straightforward WT intervals based on the variance matrices presented in Theorems \ref{thm:varCVREG} and \ref{thm:varGeo}.

We consider results for two factor moderators (allocation and ethnicity) and two numeric moderators (absolute latitude and standard quality of care). In these simulations we found that using the REML estimator of $\tau^2$ performed best. We performed simulations for the numeric moderators, based on the initial data before replicating the data set once, twice and three times (to increase the number of studies present in the analysis). For example, the Ablat results seen in Table~\ref{tab:reg_means_final} contains 13 studies, we replicate the data once to run simulations with $k = 26$, twice for simulations with $k = 39$ studies and three times for simulations with $k = 52$ studies.

\begin{table}[ht]
\centering
\caption{Coverages calculated using 10,000 trials with various moderating variables using CV$_B$'s for both DL and REML estimators of $\tau^2$. CIs were calculated using the WT intervals. We show single level data for the two data sets \cite{colditz1994efficacy, de2009standard}. In this example we chose 20 values from between the minimum of the corresponding moderator to the maximum (not including zero). Here, $k$ represents the number of studies included in the meta-regression (by adding replications of the original data set).}
\bigskip
\begin{tabular}{llccc}
  \hline
Data & Moderator & waCV & waLCV & GM \\ 
  \hline
&Allocation & 0.920 & 0.963 & 0.956 \\ 
&  Allocation [REML] & 0.933 & 0.970 & 0.965 \\  
&    Ablat ($k = 13$) & 0.947 & 0.998 & 0.998 \\ 
&  Ablat[REML]($k = 13$) & 0.909 & 0.996 & 0.996 \\ 
\cite{colditz1994efficacy} &  Ablat ($k = 26$) & 0.926 & 0.995 & 0.995 \\ 
&  Ablat[REML]($k = 26$) & 0.908 & 0.990 & 0.990 \\ 
&  Ablat ($k = 39$) & 0.922 & 0.995 & 0.995 \\ 
&  Ablat[REML]($k = 39$) & 0.914 & 0.977 & 0.977 \\ 
&  Ablat ($k = 52$) & 0.926 & 0.989 & 0.989 \\ 
&  Ablat[REML]($k = 52$) & 0.919 & 0.968 & 0.968 \\ 
  \hline
&  Ethnicity & 0.910 & 0.965 & 0.965 \\  
&  Ethnicity [REML] & 0.920 & 0.961 & 0.961 \\ 
&  SCQ ($k = 13$) & 0.910 & 0.965 & 0.965 \\ 
&  SCQ [REML] ($k = 13$) & 0.920 & 0.961 & 0.961 \\ 
\cite{de2009standard} &  SCQ ($k = 26$) & 0.931 & 0.956 & 0.956 \\ 
&  SCQ [REML] ($k = 26$) & 0.941 & 0.955 & 0.955 \\ 
&  SCQ ($k = 39$) & 0.939 & 0.957 & 0.957 \\ 
&  SCQ [REML] ($k = 39$) & 0.946 & 0.956 & 0.956 \\ 
&  SCQ ($k = 52$) & 0.944 & 0.955 & 0.955 \\ 
&  SCQ [REML] ($k = 52$) & 0.951 & 0.957 & 0.957 \\ 
   \hline
\end{tabular}
\label{tab:reg_means_final}
\end{table}

    In Table~\ref{tab:reg_means_final} we consider the results of several simulations. Note that the coverage of $waLCV$ and $GM$ for the numeric moderators are the same because the weights are equal in these simulations. We begin by showing the coverage for a geometric mean and a weighted average for two different factor simulations (described above). We see that coverage can be quite liberal for the $waCV$ intervals though log-transformed intervals and geometric means have good coverage. Continuing on, we display the results for two sets of continuous numeric moderators. Looking at the \cite{colditz1994efficacy} results, the $waCV$ intervals show liberal coverage which increases with the number of studies. For the $waLCV$ and $GM$ intervals, coverage tends to begin conservatively and become closer to the nominal as more studies are added. The coverages for the \cite{de2009standard} data set are generally good with $waCV$ intervals tending to be liberal.

\subsection{Examples}\label{ch4examples}
In this section we will give the reader an example of how these measures may be used in practice. First, consider the data set seen in \cite{colditz1994efficacy}. Remember that this data set contains 13 studies and we use absolute latitude as a moderating variable (see Section \ref{ch4means} for more detailed information). Performing a meta-regression, we see the following: $\widehat{\beta}_0 = 0.260$, $\widehat{\beta}_1 = -0.029$ and $\widehat{\tau}^2 = 0.063$. A quick calculation provides the values of CV$_B$, $M_1$ and $M_2$, for each value of absolute latitude. We see these results in Table \ref{tab:ch4example}. 

\begin{table}[h!t]
    \centering
        \caption{Contains values for heterogeneity measures when using absolute latitude as a moderating variable. These are examples of output a researcher might receive when using these measures in practice.}
        \resizebox{\columnwidth}{!}{
    \begin{tabular}{l|ccccccccc}
    \hline
       Mods  & 13 & 18 & 19 & 27 & 33 & 42 & 44 & 52 & 55 \\
       $\beta_x$ & -0.120& -0.120& -1.027& -0.296& -0.530& -0.968& -1.027& -1.260& -1.348\\
       \hline
       CV$_B$  & 2.089 & 0.944 & 0.851 & 0.475 & 0.357 & 0.260 & 0.245 & 0.200 & 0.187 \\
       CV$_B$ [REML] & 2.178 & 1.015 & 0.917 & 0.517 & 0.390 & 0.285 & 0.269 & 0.219 & 0.205 \\ 
       $M_1$ & 0.676 & 0.486 & 0.460 & 0.322 & 0.263 & 0.206 & 0.197 & 0.166 & 0.157 \\
       $M_1$ [REML] & 0.685 & 0.504 & 0.478 & 0.341 & 0.280 & 0.222 & 0.212 & 0.180 & 0.170 \\
       \hline
    \end{tabular}}
     \begin{tablenotes}
     \item\footnotesize CIs using DL for specific moderators using PropImp intervals and a CI for the weighted average:\\
\item \footnotesize $13:$ CV$_B$ $= [0.445, >>1000]$ and $M_1 = [0.308, 1.000]$.\\
\item \footnotesize $55:$ CV$_B$ $= [0.094, 0.735]$ and $M_1 = [0.086, 0.424]$.\\
\item \footnotesize $waCV = 0.742 [0.000,  1.863]$.
 \end{tablenotes}
    \label{tab:ch4example}
\end{table}

These results tell us that heterogeneity relative to the size of the effect is larger for smaller values of absolute latitude.  Hence, effects measured at larger latitudes (e.g., 55) would be relatively stable in terms of magnitude.   However, for lower latitude values such as within the tropics, the magnitude of the effect may vary considerably to the extent that small, moderate, or large effects may be observed.  Hence, while the effects themselves may vary similarly at any given latitude, given that we are interested in magnitude of the effects then we need to consider heterogeneity relative to the size of the effect.  These results indicate to the researcher that results of the usage of this vaccine near the equator could vary considerably. In addition, further research into the causes (and subsequent fixes) of the variation in the effectiveness of the vaccine near the equator could be considered.  

\section{Conclusion}\label{ch4con}

We consider the use of the CV in meta-analysis in the context of meta-regression.  We have used the CV for fixed values of the moderator, as well as means of the CV over several moderator values.  We also considered transformed values of the CV and associated CIs as was considered by \cite{cairns2020ratio}. It is of interest to be able to assess how much adding moderators affects the amount of explained heterogeneity of a meta-analysis relative to the size of the effect. This is particularly useful when such an investigation is used in conjunction with a consideration of how much that moderator affects the uncertainty of the meta-analysis (i.e., a discussion of $I^2$) creating strong hypothesis generating discussions. CIs for $\text{CV}_B$, $M_1$ and $M_2$ show strong coverage. Simulations show that PropImp intervals for  CV$_B$, $M_1$ and $M_2$ provide reliable, slightly conservative intervals in all cases. Thus, we would recommend that in addition to $I^2$, either CV$_B$ and $M_1$ should be reported, along with corresponding CIs.

\bigskip

\noindent{\it{All data and code used in this paper are available at \url{https://osf.io/f68za/}}}

\newpage

\section*{Appendix}

\subsection*{A.1.\enspace Proof of Theorem \ref{thm:varGeo}}\label{ch4appProof1}
We will prove Theorem \ref{thm:varGeo} by calculating the $ij$-th entry. Let $\log(\widehat{\text{CV}}_i) = \log(\widehat{\tau}) - \log(|\widehat{\beta}_i|)$ and $\log(\widehat{\text{CV}}_j) = \log(\widehat{\tau}) - \log(|\widehat{\beta}_j|)$ and $\widehat{\tau} = \sqrt{\widehat{\tau}^2}$ (i.e. so that we consider $\widehat{\text{CV}}_j$ a function of $\widehat{\tau}^2$ and write $\widehat{\text{CV}}_j=\sqrt{\widehat{\tau}^2}/|\widehat{\beta}_j|$). Noting that in general $d|x|/(dx)=\text{sign}(x)$, then a first order Taylor approximation has the form
\begin{equation*}
    \log(\widehat{\text{CV}})_i \approx \log(\sqrt{\tau^2}) - \log(|\beta_i|) + (\widehat{\tau}^2 - \tau^2)\cdot\frac{1}{2\cdot\tau^2} - (\widehat{\beta}_i - \beta_i)\cdot \frac{1}{|\beta_i|}\cdot \text{ sign}(\beta_i),
\end{equation*}
where an approximate $\log(\text{CV}_j)$ can be found in the same way. Therefore the $ij$-th element of the variance-covariance matrix for two dependent $\log(\text{CV}_B)$'s has the form

\begin{align*}
    \text{Cov}\left[\log(\sqrt{\tau^2}) \right.&- \log(|\beta_i|) + (\widehat{\tau}^2 - \tau^2)\cdot\frac{1}{2\cdot\tau^2} - (\widehat{\beta}_i - \beta_i)\cdot \frac{1}{|\beta_i|}\cdot \text{ sign}(\beta_i), \\
    &\left.\log(\sqrt{\tau^2}) - \log(|\beta_j|) + (\widehat{\tau}^2 - \tau^2)\cdot\frac{1}{2\cdot\tau^2} - (\widehat{\beta}_j - \beta_j)\cdot \frac{1}{|\beta_j|}\cdot \text{ sign}(\beta_j)\right].
\end{align*}

Using the fact that $\text{Cov}(aW + bX, cY + dZ) = \text{Cov}(aW, cY) + \text{Cov}(aW, dZ) + \text{Cov}(bX, cY) + \text{Cov}(bX, dZ)$, we see that 
\begin{align*}
    \text{Cov}\left[\right.&\log(\widehat{CV}_i), \left.\log(\widehat{CV}_j)\right] \approx  \ \text{Cov}\left[(\widehat{\tau}^2 - \tau^2)\cdot\frac{1}{2\cdot\tau^2}, (\widehat{\tau}^2 - \tau^2)\cdot\frac{1}{2\cdot\tau^2}\right]\\ 
    &+ \text{Cov}\left[(\widehat{\tau}^2 - \tau^2)\cdot\frac{1}{2\cdot\tau^2}, - (\widehat{\beta}_j - \beta_j)\cdot \frac{1}{|\beta_j|}\cdot \text{ sign}(\beta_j)\right]\\
    &+ \text{Cov}\left[- (\widehat{\beta}_i - \beta_i)\cdot \frac{1}{|\beta_i|}\cdot \text{ sign}(\beta_i), (\widehat{\tau}^2 - \tau^2)\cdot\frac{1}{2\cdot\tau^2}\right]\\
    &+ \text{Cov}\left[- (\widehat{\beta}_i - \beta_i)\cdot \frac{1}{|\beta_i|}\cdot \text{ sign}(\beta_i), - (\widehat{\beta}_j - \beta_j)\cdot \frac{1}{|\beta_j|}\cdot \text{ sign}(\beta_j)\right].
\end{align*}
Since $\widehat{\tau}$ and $\widehat{\beta}$ are independent so that the second and third terms above are zero, and since E$[\widehat{\tau}^2] \approx \tau^2$ and E$[\widehat{\beta}] \approx \beta$, then by the definition of the covariance function
\begin{align*}
    \text{Cov}[&\log(\widehat{\text{CV}}_i), \log(\widehat{\text{CV}}_j)] \approx \ \text{Cov}\left[(\widehat{\tau}^2 - \tau^2)\cdot\frac{1}{2\cdot\tau^2}, (\widehat{\tau}^2 - \tau^2)\cdot\frac{1}{2\cdot\tau^2}\right]\\ 
    & + \text{Cov}\left[- (\widehat{\beta}_i - \beta_i)\cdot \frac{1}{|\beta_i|}\cdot \text{ sign}(\beta_i), - (\widehat{\beta}_j - \beta_j)\cdot \frac{1}{|\beta_j|}\cdot \text{ sign}(\beta_j)\right]\\
    &= E\left[\frac{(\tau^2 - \widehat{\tau}^2)^2}{4\cdot(\tau^2)^2}\right] + E\left[\frac{1}{|\beta_i|\cdot |\beta_j|}\cdot \text{ sign}(\beta_i)\text{ sign}(\beta_j)(\widehat{\beta_i} - \beta)(\widehat{\beta_j} - \beta)\right]\\
    &= \frac{\text{Var}(\widehat{\tau}^2)}{4\cdot\tau^4} + \frac{1}{|\beta_i|\cdot |\beta_j|}\cdot \text{ sign}(\beta_i)\text{ sign}(\beta_j)\cdot\text{Cov}(\widehat{\beta}_i, \widehat{\beta}_j).
\end{align*}
Thus a variance-covariance matrix for $k$ dependent $\log(\widehat{\text{CV}}_B)$'s is the matrix whose $ij$-th element is $$\text{Cov}[\log(\widehat{\text{CV}}_i), \log(\widehat{\text{CV}}_j)].$$  By introducing the notations just prior to Theorem \ref{thm:varGeo}, it is simple to represent this in matrix form.

\subsection*{A.2.\enspace Proof of Theorem \ref{thm:varCVREG}} \label{ch4appProof2}
Recall the methods used in Appendix A. We will prove Theorem \ref{thm:varCVREG} by calculating the $ij$-th entry. Let $\widehat{\text{CV}}_i = \frac{\widehat{\tau}}{|\widehat{\beta}_i|}$ and $\widehat{\text{CV}}_j = \frac{\widehat{\tau}}{|\widehat{\beta}_j|}$ and $\widehat{\tau} = \sqrt{\widehat{\tau}^2}$ (i.e., we consider $\widehat{\text{CV}}_j$ a function of $\widehat{\tau}^2$ and write $\widehat{\text{CV}}_j=\sqrt{\widehat{\tau}^2}/|\widehat{\beta}_j|$). Noting that in general $d|x|/(dx)=\text{sign}(x)$, then a first order Taylor approximation has the form
\begin{equation*}
    \widehat{\text{CV}}_i \approx \frac{\sqrt{\tau^2}}{|\beta_i|} + \frac{\widehat{\tau}^2 - \tau^2}{|\beta_i|}\cdot\frac{1}{2\sqrt{\tau^2}} - (\widehat{\beta}_i - \beta_i)\cdot \frac{\sqrt{\tau^2}}{\beta_i^2}\cdot \text{ sign}(\beta_i),
    \label{vcvCV}
\end{equation*}
where an approximate $\text{CV}_j$ can be found in the same way. Therefore the $ij$-th element of the variance-covariance matrix for two dependent $\text{CV}_B$'s has the form
\begin{align*}
\text{Cov}\left[\frac{\sqrt{\tau}^2}{|\beta_i|}\right.& + \frac{\widehat{\tau}^2 - \tau^2}{|\beta_i|}\cdot\frac{1}{2\sqrt{\tau^2}} - (\widehat{\beta}_i - \beta_i)\cdot \frac{\sqrt{\tau^2}}{\beta_i^2}\cdot \text{ sign}(\beta_i),\\ &\left.\frac{\sqrt{\tau^2}}{|\beta_j|} + \frac{\widehat{\tau}^2 - \tau^2}{|\beta_j|}\cdot\frac{1}{2\sqrt{\tau^2}} - (\widehat{\beta}_j - \beta_j)\cdot \frac{\sqrt{\tau^2}}{\beta_j^2}\cdot \text{ sign}(\beta_j)\right].
\end{align*}
Using the fact that $\text{Cov}(aW + bX, cY + dZ) = \text{Cov}(aW, cY) + \text{Cov}(aW, dZ) + \text{Cov}(bX, cY) + \text{Cov}(bX, dZ)$, we see that 

\begin{align*}
    \text{Cov}(&\widehat{\text{CV}}_i, \widehat{\text{CV}}_j) \approx \ \text{Cov}\left[\frac{\widehat{\tau}^2 - \tau^2}{|\beta_i|}\right.\cdot\frac{1}{2\sqrt{\tau^2}} - (\widehat{\beta}_i - \beta_i)\cdot \frac{\sqrt{\tau^2}}{\beta_i^2}\cdot \text{ sign}(\beta_i), \\
    &\qquad\qquad\qquad\qquad\quad\left.\frac{\widehat{\tau}^2 - \tau^2}{|\beta_j|}\cdot\frac{1}{2\sqrt{\tau^2}} - (\widehat{\beta}_j - \beta_j)\cdot \frac{\sqrt{\tau^2}}{\beta_j^2}\cdot \text{ sign}(\beta_j)\right]\\
    &= \text{Cov}\left[\frac{\widehat{\tau}^2 - \tau^2}{|\beta_i|}\cdot\frac{1}{2\sqrt{\tau^2}}, \frac{\widehat{\tau}^2 - \tau^2}{|\beta_j|}\cdot\frac{1}{2\sqrt{\tau^2}}\right] \\
    &+ \text{Cov}\left[\frac{\widehat{\tau}^2 - \tau^2}{|\beta_i|}\cdot\frac{1}{2\sqrt{\tau^2}}, - (\widehat{\beta}_j - \beta_j)\cdot \frac{\sqrt{\tau^2}}{\beta_j^2}\cdot \text{ sign}(\beta_j)\right]\\
    &+ \text{Cov}\left[- (\widehat{\beta}_i - \beta_i)\cdot \frac{\sqrt{\tau^2}}{\beta_i^2}\cdot \text{ sign}(\beta_i), \frac{\widehat{\tau}^2 - \tau^2}{|\beta_j|}\cdot\frac{1}{2\sqrt{\tau^2}}\right]\\
    &+ \text{Cov}\left[- (\widehat{\beta}_i - \beta_i)\cdot \frac{\sqrt{\tau^2}}{\beta_i^2}\cdot \text{ sign}(\beta_i), - (\widehat{\beta}_j - \beta_j)\cdot \frac{\sqrt{\tau^2}}{\beta_j^2}\cdot \text{ sign}(\beta_j)\right].\\
\end{align*}
Thus, remembering from Appendix 4A that $\widehat{\tau}$ and $\widehat{\beta}$ are independent so that the second and third terms above are zero, and E$[\widehat{\tau}^2] \approx \tau^2$ and E$[\widehat{\beta}] \approx \beta$, then by the definition of the covariance function
\begin{align*}
    \text{Cov}(&\widehat{\text{CV}}_i, \widehat{\text{CV}}_j) \approx \ \text{Cov}\left[\frac{\widehat{\tau}^2 - \tau^2}{|\beta_i|}\cdot\frac{1}{2\sqrt{\tau^2}}, \frac{\widehat{\tau}^2 - \tau^2}{|\beta_j|}\cdot\frac{1}{2\sqrt{\tau^2}}\right]\\ 
    &\qquad\qquad\quad + \text{Cov}\left[- (\widehat{\beta}_i - \beta_i)\cdot \frac{\sqrt{\tau^2}}{\beta_i^2}\cdot \text{ sign}(\beta_i), - (\widehat{\beta}_j - \beta_j)\cdot \frac{\sqrt{\tau^2}}{\beta_j^2}\cdot \text{ sign}(\beta_j)\right]\\
    &= E\left[\frac{(\tau^2 - \widehat{\tau}^2)^2}{4\cdot\tau^2\cdot|\beta_i||\beta_j|}\right] + E\left[\frac{\tau^2}{\beta_i^2\cdot \beta_j^2}\cdot \text{ sign}(\beta_i)\text{ sign}(\beta_j)(\widehat{\beta_i} - \beta)(\widehat{\beta_j} - \beta)\right]\\
    &= \frac{\text{Var}(\widehat{\tau}^2)}{4\cdot\tau^2\cdot|\beta_i||\beta_j|} + \frac{\tau^2}{\beta_i^2\cdot \beta_j^2}\cdot \text{ sign}(\beta_i)\text{ sign}(\beta_j)\cdot\text{Cov}(\widehat{\beta}_i, \widehat{\beta}_j).
\end{align*}
Thus a variance-covariance matrix for $k$ dependent $\widehat{\text{CV}}_B$'s is the matrix whose $ij$-th element is $\text{Cov}(\widehat{\text{CV}}_i, \widehat{\text{CV}}_j)$. By introducing the notations just prior to Theorem \ref{thm:varGeo}, it is simple to represent this in matrix form.
\newpage
\bibliographystyle{authordate4}
\bibliography{main}

\begin{thebibliography}{}

\bibitem[\protect\citename{Baker {\em et~al.}, }2009]{baker2009understanding}
{\sc Baker, W.~L., Michael~White, C., Cappelleri, J.~C., Kluger, J., Coleman,
  C.~I., \& {From the Health Outcomes, Policy, and Economics (HOPE)
  Collaborative Group}}. 2009.
\newblock Understanding heterogeneity in meta-analysis: {T}he role of
  meta-regression.
\newblock {\em International {J}ournal of {C}linical {P}ractice}, {\bf 63}(10),
  1426--1434.

\bibitem[\protect\citename{Borenstein {\em et~al.},
  }2011]{borenstein2011introduction}
{\sc Borenstein, M., Hedges, L.~V., Higgins, J. P.~T., \& Rothstein, H.~R.}
  2011.
\newblock {\em Introduction to meta-analysis}.
\newblock John Wiley \& Sons.

\bibitem[\protect\citename{Borenstein {\em et~al.},
  }2017]{borenstein2017basics}
{\sc Borenstein, M., Higgins, J. P.~T., Hedges, L.~V., \& Rothstein, H.~R.}
  2017.
\newblock {Basics of meta-analysis: I2 is not an absolute measure of
  heterogeneity}.
\newblock {\em {Research Synthesis Methods}}, {\bf 8}(1), 5--18.

\bibitem[\protect\citename{Cairns \& Prendergast, }2020]{cairns2020ratio}
{\sc Cairns, M., \& Prendergast, L.~A.} 2020.
\newblock On ratio measures of heterogeneity for meta-analyses.
\newblock {\em {Research Synthesis Methods}}.
\newblock \url{https://onlinelibrary.wiley.com/doi/10.1002/jrsm.1517}.

\bibitem[\protect\citename{Card, }2015]{card2015applied}
{\sc Card, N.~A.} 2015.
\newblock {\em Applied meta-analysis for social science research}.
\newblock Guilford Publications.

\bibitem[\protect\citename{Colditz {\em et~al.}, }1994]{colditz1994efficacy}
{\sc Colditz, G.~A., Brewer, T.~F., Berkey, C.~S., Wilson, M.~E., Burdick, E.,
  Fineberg, H.~V., \& Mosteller, F.} 1994.
\newblock Efficacy of {BCG} vaccine in the prevention of tuberculosis:
  {M}eta-analysis of the published literature.
\newblock {\em {JAMA}}, {\bf 271}(9), 698--702.

\bibitem[\protect\citename{de~Bruin {\em et~al.}, }2009]{de2009standard}
{\sc de~Bruin, M., Viechtbauer, W., Hospers, H.~J., Schaalma, H.~P., \& Kok,
  G.} 2009.
\newblock Standard care quality determines treatment outcomes in control groups
  of {HAART}-adherence intervention studies: {I}mplications for the
  interpretation and comparison of intervention effects.
\newblock {\em Health {P}sychology}, {\bf 28}(6), 668.

\bibitem[\protect\citename{DerSimonian \& Laird, }1986]{dersimonian1986meta}
{\sc DerSimonian, R., \& Laird, N.} 1986.
\newblock Meta-analysis in clinical trials.
\newblock {\em Controlled {C}linical {T}rials}, {\bf 7}, 177--188.

\bibitem[\protect\citename{Higgins \& Thompson, }2002]{higgins2002quantifying}
{\sc Higgins, J. P.~T., \& Thompson, S.~G.} 2002.
\newblock Quantifying heterogeneity in a meta-analysis.
\newblock {\em Statistics in {M}edicine}, {\bf 21}, 1539--1558.

\bibitem[\protect\citename{Higgins \& Thompson, }2004]{higgins2004controlling}
{\sc Higgins, J. P.~T., \& Thompson, S.~G.} 2004.
\newblock Controlling the risk of spurious findings from meta-regression.
\newblock {\em Statistics in {M}edicine}, {\bf 23}(11), 1663--1682.

\bibitem[\protect\citename{Langan {\em et~al.}, }2019]{langan2019comparison}
{\sc Langan, D., Higgins, J. P.~T., Jackson, D., Bowden, J., Veroniki, A.~A.,
  Kontopantelis, E., Viechtbauer, W., \& Simmonds, M.} 2019.
\newblock A comparison of heterogeneity variance estimators in simulated
  random-effects meta-analyses.
\newblock {\em Research {S}ynthesis {M}ethods}, {\bf 10}(1), 83--98.

\bibitem[\protect\citename{Lloyd, }1990]{lloyd1990confidence}
{\sc Lloyd, C.~J.} 1990.
\newblock Confidence intervals from the difference between two correlated
  proportions.
\newblock {\em {Journal of the American Statistical Association}}, {\bf
  85}(412), 1154--1158.

\bibitem[\protect\citename{Newcombe, }2011]{newcombe2011propagating}
{\sc Newcombe, R.~G.} 2011.
\newblock Propagating imprecision: {C}ombining confidence intervals from
  independent sources.
\newblock {\em {Communications in Statistics – Theory and Methods}}, {\bf
  40}(17), 3154--3180.

\bibitem[\protect\citename{{R {C}ore {T}eam}, }2020]{R}
{\sc {R {C}ore {T}eam}}. 2020.
\newblock {\em {R: A Language and Environment for Statistical Computing}}.

\bibitem[\protect\citename{Sidik \& Jonkman, }2005]{sidik2005simple}
{\sc Sidik, K., \& Jonkman, J.~N.} 2005.
\newblock Simple heterogeneity variance estimation for meta-analysis.
\newblock {\em {Journal of the Royal Statistical Society: Series C (Applied
  Statistics)}}, {\bf 54}(2), 367--384.

\bibitem[\protect\citename{Takkouche {\em et~al.},
  }1999]{takkouche1999evaluation}
{\sc Takkouche, B., Cadarso-Suarez, C., \& Spiegelman, D.} 1999.
\newblock Evaluation of old and new tests of heterogeneity in epidemiologic
  meta-analysis.
\newblock {\em American {J}ournal of {E}pidemiology}, {\bf 150}(2), 206--215.

\bibitem[\protect\citename{Takkouche {\em et~al.},
  }2013]{takkouche2013confidence}
{\sc Takkouche, B., Khudyakov, P., Costa-Bouzas, J., \& Spiegelman, D.} 2013.
\newblock Confidence intervals for heterogeneity measures in meta-analysis.
\newblock {\em {American {J}ournal of {E}pidemiology}}, {\bf 178}(6),
  993--1004.

\bibitem[\protect\citename{Thompson \& Higgins, }2002]{thompson2002should}
{\sc Thompson, S.~G., \& Higgins, J. P.~T.} 2002.
\newblock How should meta-regression analyses be undertaken and interpreted?
\newblock {\em Statistics in {M}edicine}, {\bf 21}(11), 1559--1573.

\bibitem[\protect\citename{Veroniki {\em et~al.}, }2016]{veroniki2016methods}
{\sc Veroniki, A.~A., Jackson, D., Viechtbauer, W., Bender, R., Bowden, J.,
  Knapp, G., Kuss, O., Higgins, J. P.~T., Langan, D., \& Salanti, G.} 2016.
\newblock Methods to estimate the between-study variance and its uncertainty in
  meta-analysis.
\newblock {\em Research {S}ynthesis {M}ethods}, {\bf 7}, 55--79.

\bibitem[\protect\citename{Viechtbauer, }2010]{metafor}
{\sc Viechtbauer, W.} 2010.
\newblock Conducting meta-analyses in {R} with the {metafor} package.
\newblock {\em Journal of {S}tatistical {S}oftware}, {\bf 36}(3), 1--48.

\end{thebibliography}
\end{document}